\documentclass[]{aastex}

\usepackage{emulateapj5}
\usepackage{onecolfloat}
\usepackage{graphicx} 
\usepackage{fancyheadings} 
\usepackage{ulem}
\usepackage{rotating}
\usepackage{lscape}

\newcommand{\vlim}{v_{lim}}

\newcommand{\delv}{\Delta v}

\newcommand{\lya}{Ly$\alpha$ }

\newcommand{\kms}{km~s$^{-1}$ }
\newcommand{\cm}[1]{\, {\rm cm^{#1}}}
\newcommand{\N}[1]{{N({\rm #1})}}

\newcommand{\mkms}{{\rm \; km\;s^{-1}}}

\begin{document}

\twocolumn[%
\submitted{Submitted to the Astrophysical Journal Letters: August 7, 2001}

\title{ARE SIMULATIONS OF CDM 
CONSISTENT WITH GALACTIC-SCALE OBSERVATIONS AT HIGH REDSHIFT?}

\author{ JASON X. PROCHASKA\altaffilmark{1}}
\affil{The Observatories of the Carnegie Institute of Washington}
\affil{813 Santa Barbara St. \\
Pasadena, CA 91101}
\email{xavier@ociw.edu}
\and
\author{ARTHUR M. WOLFE\altaffilmark{1}}
\affil{Department of Physics, and Center for Astrophysics and Space Sciences}
\affil{University of California, San Diego; 
C--0424; La Jolla, CA 92093}
\email{awolfe@ucsd.edu}

\begin{abstract} 

We compare new observations on the kinematic characteristics of 
the damped \lya systems against results from numerical 
SPH simulations to test the predictions of hierarchical galaxy
formation.  This exercise is particularly motivated by recent 
numerical results on the cross-section of damped \lya systems.
Our analysis focuses on the velocity
widths $\delv$ of $\approx 50$ low-ion absorption profiles from our
sample of $z>1.5$ damped \lya systems.  
The results indicate that current 
numerical simulations fail to match the damped \lya observations
at high confidence levels ($>99.9 \%$).
Although we do not believe that
our results present an insurmountable challenge to the paradigm of
hierarchical cosmology, the damped \lya observations suggest that
current numerical SPH simulations overlook an integral aspect
of galaxy formation.


\end{abstract}

]
\altaffiltext{1}{Visiting Astronomer, W.M. Keck Telescope.
The Keck Observatory is a joint facility of the University
of California and the California Institute of Technology.}

\pagestyle{fancyplain}
\lhead[\fancyplain{}{\thepage}]{\fancyplain{}{PROCHASKA \& WOLFE}}
\rhead[\fancyplain{}{ARE CDM SIMULATIONS 
CONSISTENT WITH OBSERVATIONS AT HIGH $z$?}
]{\fancyplain{}{\thepage}}
\setlength{\headrulewidth=0pt}
\cfoot{}

\section{INTRODUCTION}

With the apparent agreement between recent CMB experiments 
\citep{boom01,maxima01,dasi01}, 
high redshift supernovae surveys \citep{perlmutter99,riess01},
and several other cosmological diagnostics 
\citep[e.g. cluster surveys,][]{bahcall99},
there is growing optimism that adiabatic Cold Dark Matter (CDM) 
cosmology accurately describes our universe on linear scales.  
Over the past two decades researchers
have built a theoretical framework for evolving the
initial adiabatic fluctuations to the non-linear regime 
\citep[e.g.][]{peebles80}.
To examine the formation and evolution of galaxies, researchers
have introduced semi-analytic formalism \citep[e.g.][]{press74} and 
pursued comprehensive numerical simulations \citep[e.g.][]{davis85}. 
These galaxy formation scenarios 
are constrained and tested against the tremendous
database of galactic observations at low redshift. 
The models have achieved reasonable success in matching the 
luminosity functions, number counts, correlation functions, and 
Tully-Fisher relation of low redshift galaxies. 
On the other hand, there are still
serious inconsistencies between theoretical prediction and
observation on small scales including the
inner density profiles of galaxies \citep[e.g.][]{deblok01}
and the angular momentum of spiral disks \citep[e.g.][]{navarro00}.
Nevertheless,  this research presents the community with important
insight into the processes of galaxy formation.

Unfortunately, there are few observations on galactic scales at high
redshift and therefore few constraints on models of protogalaxies within
hierarchical cosmology.
The most significant population of high $z$ protogalaxies identified
in emission are the Lyman break galaxies \citep{ste96}.
These rapidly star forming protogalaxies -- 
selected with broad band filters designed to identify a strong decrement
in flux below the Lyman break -- have provided significant
insight into star formation in the early universe.
Observations of the Lyman break clustering properties \citep{giav98,adelb98}
and luminosity function \citep{std99} have provided measurements on the
protogalactic star formation rate at $z>3$ \citep{steid98,madau96}, 
insight into the sub-structure of protogalactic halos 
\citep{weschsler98,bullock01}, and hints at chemical evolution
\citep{pettini01}.
Another active area of high $z$ 
protogalactic research derives from the spectral analysis
of distant quasars, i.e., quasar absorption line studies.  
The damped \lya systems, \lya absorption line features with HI column 
densities in excess of 10$^{20.3} \cm{-2}$,  are widely believed
to be associated with protogalaxies in the early universe \citep{wol86}.
Because these systems are identified in absorption (i.e. weighted by 
optical depth), they should
probe protogalaxies with a broad distribution of 
physical characteristics (e.g. mass,
luminosity, morphology).    In turn, the analysis of high resolution spectroscopy
allows one to explore the chemical abundances 
\citep{lu96,pw99,pw01}, metallicity evolution 
\citep{ptt94,ptt97,pw00,pgw01}, star formation rate \citep{wol01},
and kinematic properties (Prochaska \& Wolfe 1997, hereafter PW97;
Prochaska \& Wolfe 1998, hereafter PW98; Wolfe \& Prochaska 2000a, hereafter
WP00; Wolfe \& Prochaska 2000b). 
These observations help describe the processes of
galaxy formation at early epochs and establish the
starting point from which modern galaxies evolved.

In PW97, we examined the kinematic characteristics of 17 damped \lya
systems and tested several protogalactic models against our observations.  
Of greatest importance, we tested the predictions from the leading
galaxy formation model of CDM 
which characterized protogalaxies as centrifugally supported disks
within dark matter halos \citep[e.g.][]{kauff96,mmw98}.
We found that this scenario could be ruled out at very high 
confidence level because the CDM mass spectrum at $z \sim 2$
predicts far too many galaxies with circular velocity $v_c < 100 \mkms$
to reproduce the observed
velocity widths $\delv$ of the damped \lya metal-lines.
Inspired by our results, 
Haehnelt, Steinmetz, \& Rauch (1998; hereafter H98)
examined high resolution
($\approx$ 1 kpc) numerical simulations of a small sample of dark
matter halos to investigate protogalactic 
velocity fields in the CDM cosmology.
In contrast to the semi-analytic models which assumed a rotating
disk morphology, 
H98 found the kinematics of protogalaxies were
best described by a combination of rotation, infall, and
random motions.  Therefore, the typical velocity
width in these protogalactic halos is $\delv \approx \frac{2}{3} v_c$
instead of the $\approx \frac{1}{3} v_c$ predicted by a thick,
rotating disk.  Doubling the average $\delv$ value brought
the predictions of CDM into reasonable agreement 
with the damped \lya observations 
provided the cross-section of protogalaxies scales like 
$\sigma(v_c) \propto v_c^{2.5}$ \citep{hae00}.  
This latter assumption did not arise from the numerical simulations
but followed from a simple scaling argument between radial scale-length,
$v_c$, and the cross-section.   However, a more recent analysis
of the cross-section
of damped \lya systems in a series of SPH numerical simulations indicates
$\sigma(v_c) \propto v_c^{1.25}$ \citep[][hereafter G01]{gardner01}.
In this Letter we test the implications of these numerical simulations
against new observations on the kinematic characteristics of high
redshift damped \lya systems.  
With the addition of nearly 25 systems since PW98, the damped \lya systems
place very strong constraints on the processes of galaxy formation in the
early universe.

\section{ANALYSIS}
\label{sec-anly}

\begin{figure}[hb]
\includegraphics[height=5.4in, width=3.8in]{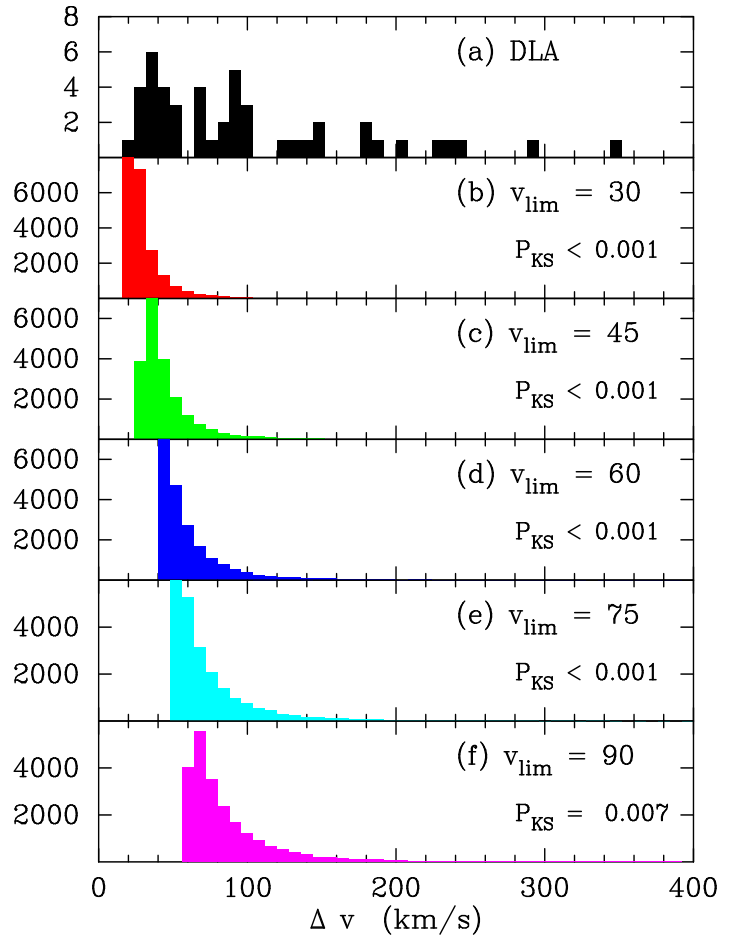}
\caption{Velocity width ($\delv$) distributions for (a) our sample of
$z>1.5$ damped \lya systems compared against the predictions from SPH
numerical simulations of the $\Lambda$CDM cosmology.  For the model
predictions, we randomly draw 500 $\delv$ values at $z_{abs}$ of each
damped \lya system according to the prescription outlined in the text.
This includes the assumption that no halos with $v_c < \vlim$ give rise
to damped \lya systems and we consider a series of $\vlim$ values:
(b) 30 km/s, (c) 45 km/s, (d) 60 km/s, (e) 75 km/s, and (f) 90 km/s.
For $\vlim < 90$~km/s the models are ruled out at very high confidence
level by the Kolmogorov-Smirnov test.
}
\label{fig:delv}
\end{figure}

In terms of the kinematic characteristics of the damped \lya systems,
the velocity width $\delv$ of the low-ion profiles\footnote{Metal-line transitions
prominent in HI regions, e.g., Si~II 1808} places the most 
fundamental constraint on protogalactic models.  This statistic
probes both the gravitational mass of the protogalaxy as
well as the specific morphology of the neutral baryonic gas.  Following
PW97, we define the $\delv$ value as the interval encompassing $90\%$
of the total optical depth.  Provided the analysis is restricted
to unsaturated low-ion transitions, this statistic is a robust  
measure of the velocity fields probed by the damped \lya sightlines.
In Figure~\ref{fig:delv}a we present
the $\delv$ values from the $z_{abs}>1.5$ damped \lya systems presented
in PW97, PW98, WP00, 13 new systems taken from \cite{pro01}, and
two systems kindly provided by \cite{ellison00} and \cite{dessauges00}.  
The values range from $\delv \approx 20 - 350 \mkms$
with a median of $\approx$85 km/s including a significant number of systems
with $\delv > 120 \mkms$.  
In a future paper \citep{pw02}, we discuss the kinematic properties
of the new systems at greater length and pursue other aspects of these
kinematic observations. 

Now consider the predictions of the damped \lya
$\delv$ distribution derived from numerical SPH simulations
of the CDM cosmology. 
Because of the competing expense of spatial resolution
and cosmological volume,  we must
combine the results of two numerical
simulations: (1) the high resolution, small volume analysis of
H98 which demonstrated $\delv \approx \frac{2}{3} v_c$ and (2) the
cross-section analysis of a cosmologically relevant volume by
G01 which showed $\sigma(v_c) \propto v_c^\beta$ with $\beta \approx 1-1.5$.
We can convolve these two results with the mass spectrum
$n_{DM}(v_c)$ of the dark matter halos at the redshift of each damped
\lya system.  In the following, we calculate $n_{DM}(v_c)$ following
the prescription of \cite{jenkins01} and limit ourselves to the
$\Lambda$CDM cosmology 
($\sigma_8 = 1.0, h=0.7, \Omega_m = 0.3, \Lambda = 0.7$).  
The most important free parameter in the analysis is the limiting velocity $\vlim$
below which a DM halo will not give rise to a damped \lya system.
The analysis by G01 suggests a value $\vlim \approx 60 \mkms$ 
is required to match 
the observed number density of damped \lya systems $n_{DLA}(z)$.   
Other analyses, however,
have suggested that halos with $v_c$ as low as 35~\kms will maintain
significant reservoirs of neutral gas \citep{thoul96,nav97,kepner97}.  
In the following we consider
a range of $\vlim$ values and stress the importance of this single parameter on
our conclusions for the damped \lya systems.  

\begin{figure}[ht]
\includegraphics[height=3.7in, width=3.2in,angle=-90]{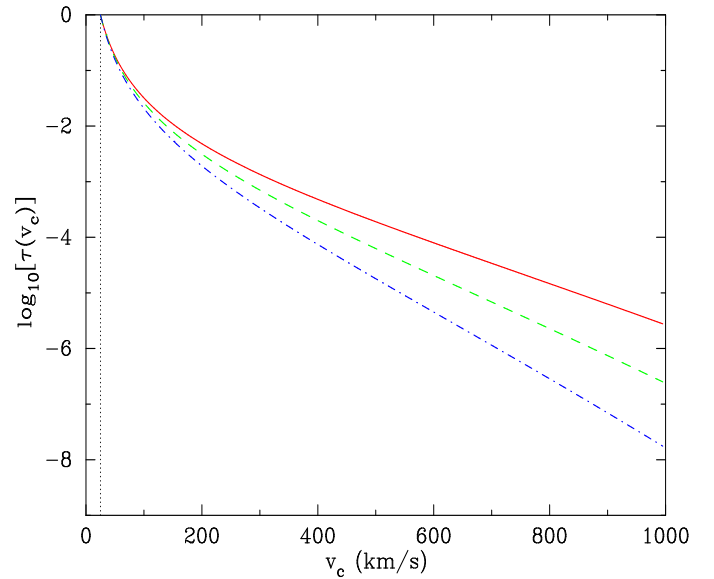}
\caption{Log $\tau(v_c)$ for damped \lya systems at redshift
$z = $ 2 (solid), 3 (dashed) and 4 (dash-dot).  The dotted line
demarcates the $v_{lim}$ value below which a dark matter halo is
presumed incapable of giving rise to a damped \lya system.
The curves demonstrate
the predicted dominance of low mass halos in the $\Lambda$CDM cosmology
for the damped \lya systems.  One also notes the evolution toward more
massive systems at lower redshift.
}
\label{fig:tau}
\end{figure}

We build the theoretical $\delv$ distributions by randomly drawing
500 dark matter halos at the absorption redshift of each 
damped \lya system weighted by optical depth 
$\tau(z, v_c) = n_{DLA}(z,v_c) \, \sigma(z, v_c)$ and assuming 
$\delv = \frac{2}{3} v_c$.   
For $\sigma(z,v_c) \propto v_c^\beta$,
we interpolate the $\beta$ values presented by G01 at $z = 2, 3$, and 4 for
the $\Lambda$CDM cosmology. Figure~\ref{fig:tau} plots
$\log \tau(z,v_c)$ at $z = 2$ (solid), 3 (dashed), and 4 (dash-dot) for
$v_c = 25 - 1000 \mkms$, where each curve has been  
normalized to have a maximum optical depth of unity.  The curves reinforce
the fundamental tenant of hierarchical cosmology: the mass spectrum at high
redshift is dominated by low mass halos which merge in time to form
more massive structures.
Furthermore, the curves in Figure~\ref{fig:tau}
emphasize the importance of the $\vlim$ parameter.  The optical 
depth will be dominated by galaxies with $v_c \approx \vlim$ 
and therefore all predictions on the kinematic
characteristics of the damped \lya systems are sensitive to this
parameter.

Figure~\ref{fig:delv} presents the $\delv$ distributions for the
$\Lambda$CDM cosmology for $\vlim = $ (b) 30 km/s, (c) 45 km/s,
(d) 60 km/s, (e) 75 km/s, and (f) 90 km/s.
For each theoretical distribution we performed a two-sided Kolmogorov-Smirnov
test against the observed damped \lya $\delv$ distribution. 
The probability $P_{KS}$ that the two distributions could have
been drawn for the same parent population is indicated in Figure~\ref{fig:delv}.
This statistic is most sensitive to differences in 
the medians of two distributions and is 
a conservative measure; low $P_{KS}$ values are indicative of qualitatively
poor matches.  The results presented in Figure~\ref{fig:delv} indicate that
models with $v_{lim} < 90 \mkms$ are ruled out at $> 99.9\%$ c.l.
This suggests a serious failure in our current understanding of
galaxy formation within the CDM cosmology.  

%[fedg discussion]
%Theoretically the results presented in Figure~\ref{fig:edg} do not pose
%a significant problem.  It may be far more likely for a sightline to
%intersect 2 or 3 clumps than five.  Observationally, however, the results
%are more worrisome.  In PW97, we demonstrated that even profiles with modest
%$\delv$ values can be comprised of $N > 10$ individual 'clouds'.  We expect
%this is a universal characteristics of the damped \lya systems.  In order
%to explain the observations while limiting the number of protogalactic
%clumps to only a few, one must allow for significant
%substructure within each protogalactic clump, substructure which currently
%lies beneath the resolution limits of numerical simulations.

\section{DISCUSSION}
\label{sec-discuss}

The damped \lya systems present a distinct challenge to models of galaxy
formation within adiabatic cosmology.  In particular, the observations
impose two well-defined, competing constraints on galaxy formation scenarios:
(1) the mass spectrum must include low enough mass halos to match the
observed number density of damped \lya systems, $n_{DLA}(z)$, while
(2) the $\delv$ distribution limits the number of low mass
halos which can contribute to damped systems.
The results presented in the previous section indicate that current
results from numerical SPH simulations cannot satisfy both constraints.
Although a model with $\vlim \approx 100 \mkms$ could reproduce the
median of the observed $\delv$ distribution, this scenario would
severely underpredict $n_{DLA}$.
Therefore, physical processes which predict large $v_{lim}$ values
without modifying the cross-sectional dependence of the dark matter halos
cannot simultaneously satisfy the two constraints. 
At present, we do not consider this 
failure to be a fatal blow to the paradigm of 
hierarchical galaxy formation, but it does require that at least
one of our theoretical conditions is incorrect.

One possibility is that G01 incorrectly determined the cross-sectional
dependence of dark matter halos containing damped \lya systems.
In their analysis, G01 had to extrapolate 
the results from their SPH simulations
to mass scales beneath their numerical resolution.
The authors argued that higher resolution simulations would be unlikely to 
modify their results, but processes on sub-kpc scales
(e.g. supernovae feedback, tidal gas stripping) could have profound
implications for the physical nature of the damped \lya
systems.  If such processes implied significantly larger cross-sections
for more massive halos, then the predicted $\delv$ distribution 
would better match the observed distribution.  
To maintain an accurate prediction for $n_{DLA}$, however,
the simulations would require a $\vlim$ value even greater than the
$\approx 60 \mkms$ implied by G01.  Because this value already exceeds
estimates due to the effects of photoionization \citep{nav97,kepner97,gnedin00},
one would need to introduce other physical processes to eliminate neutral
gas from within low mass halos.
Perhaps supernovae feedback also plays a significant role here.

Another possible solution is that the $\delv \approx \frac{2}{3} v_c$
relation suggested by H98 is incorrect, particularly in the
low $v_c$ regime.  If physical processes like supernovae feedback
doubled the typical velocity width, then the simulations would match
the median of the observed $\delv$ distribution 
with $v_{lim} \approx 50 \mkms$.  At the same time, however,
the theoretical distribution would significantly underpredict the
number of damped \lya systems with $\delv > 120 \mkms$ because
the optical depth is so heavily weighted to low $v_c$ halos.
Furthermore, a model driven by non-gravitational processes would
need to be finely tuned to reproduce the line-profile asymmetries and
the correlations observed for the low-ion and C~IV gas in damped
\lya systems (WP00).  These issues could be accurately addressed in
a relatively inexpensive simulation similar to H98 but including the
effects of supernovae feedback.
We stress that if feedback processes are central to an understanding
of the damped \lya systems then they are likely to dramatically impact
the formation and evolution of all galaxies.

The only remaining ingredient of our treatment is the mass spectrum
of dark matter halos.  
In this Letter, we have adopted the \cite{jenkins01} formalism
which is an analytic fit to the results from the Hubble Volume 
simulations \citep{evrard01}.
This formalism -- at least in part -- avoids the overestimate of
low mass halos identified in 
the Press-Schecter formalism.
We warn, however, that our analysis considers dark matter halos 
below the mass resolution of the simulations analysed
by \cite{jenkins01}.  
It is possible that even this analytic formalism overpredicts
the mass density of these low mass halos but we consider it unlikely that
the difference would be large enough for the simulations to match
the $\delv$ distribution with $v_{lim} < 80 \mkms$.

In contrast to the numerical simulations, semi-analytic models of
galaxy formation can simultaneously account for the constraints
imposed by the damped \lya observations. 
As described by \cite{maller01},
the $\delv$ distribution of the damped \lya systems can be
reproduced by a scenario similar to H98 provided a significant
fraction of the satellites residing in dark matter 
halos have large gaseous extent (i.e. cross-sectional dependence 
$\sigma(v_c) \propto v_c^{2.5}$).
There are difficulties associated with this scenario, however.  
For example, these satellites have very low surface density which leads
to a prediction for the $\N{HI}$ frequency distribution, $f(N)$, considerably
steeper than observed.  Furthermore, the gas distributions
adopted by \cite{maller01} were driven by the observations and may be 
difficult to motivate physically, particularly if the numerical simulations
indicate very different gas profiles.  Nevertheless, the semi-analytic
models may be revealing the path that the numerical simulations must
ultimately travel; indeed, this is the strength of semi-analytic
modeling.

In closing, we stress the need for a simulation
with large cosmological volume (e.g. $> 20 h^{-1}~$Mpc)
and sufficiently high spatial resolution to self-consistently examine
the cross-sectional dependence of the damped \lya systems and probe the
underlying velocity fields.  If the results from H98 and G01 stand, the
damped \lya observations demonstrate that processes critical to the 
galaxy formation have been overlooked by current scenarios.

\acknowledgments

We thank E. Gawiser for comments and criticism. 
J.X.P. acknowledges support from a
Carnegie postdoctoral fellowship and thanks CITA for their hospitality
while this paper was written. AMW was partially supported by 
NSF grant AST 0071257.

\end{document}